\documentclass{KapProc}
\usepackage{graphicx}
\kluwerbib

\begin{document}
\def\farcs{\hbox{$.\!\!^{\prime\prime}$}}
\def\la{\mathrel{\mathchoice {\vcenter{\offinterlineskip\halign{\hfil
$\displaystyle##$\hfil\cr<\cr\sim\cr}}}
{\vcenter{\offinterlineskip\halign{\hfil$\textstyle##$\hfil\cr
<\cr\sim\cr}}}
{\vcenter{\offinterlineskip\halign{\hfil$\scriptstyle##$\hfil\cr
<\cr\sim\cr}}}
{\vcenter{\offinterlineskip\halign{\hfil$\scriptscriptstyle##$\hfil\cr
<\cr\sim\cr}}}}}
\def\ga{\mathrel{\mathchoice {\vcenter{\offinterlineskip\halign{\hfil
$\displaystyle##$\hfil\cr>\cr\sim\cr}}}
{\vcenter{\offinterlineskip\halign{\hfil$\textstyle##$\hfil\cr
>\cr\sim\cr}}}
{\vcenter{\offinterlineskip\halign{\hfil$\scriptstyle##$\hfil\cr
>\cr\sim\cr}}}
{\vcenter{\offinterlineskip\halign{\hfil$\scriptscriptstyle##$\hfil\cr
>\cr\sim\cr}}}}}

\renewcommand{\floatpagefraction}{0.7}
\newcommand{\subs}[1]{\mbox{\scriptsize #1}}

\articletitle{The luminosity function of QSO host galaxies} 

\author{
  Lutz Wisotzki$^1$, Bj\"orn Kuhlbrodt$^2$, Knud Jahnke$^2$}
\affil{$^1$Universit\"at Potsdam,
       $^2$Hamburger Sternwarte}
\email{$^1$lutz@astro.physik.uni-potsdam.de}

\begin{abstract}
We report on results from $H$ band imaging observations 
of a complete sample of high-luminosity low-redshift QSOs.
The luminosity function of QSO hosts is similar
in shape to that of normal galaxies, although offset 
in normalisation by a factor of $10^{-4}$. 
This supports the hypothesis that 
the parent population of quasars is identical to 
the general population of early-type field galaxies.
\end{abstract}

\section{Introduction}

Identifying the parent population of QSO host galaxies
is one of the fundamental problems linking the QSO phenomenon
to galaxy evolution in general. While the most widely adopted
approach is to compare morphological (and increasingly
also spectral) properties of normal and active galaxies,
we investigate here the statistical distribution of galaxy
luminosities. This short report highlights some results
which will be elaborated in more detail in a series of 
forthcoming papers. We adopt $h = 0.5$ and $q_0 = 0.5$
throughout.

\section{\textit{H} band imaging of a complete QSO sample}

In order to constrain the overall luminosity distribution of QSO hosts,
the investigated targets must represent a \emph{fair sample} of the 
QSO population. To obtain also the normalisation, the sample has 
furthermore to be \emph{complete} (in the sense of comprising all 
QSOs within a given area that are brighter than a well-defined flux limit).
Our sample has been selected from the Hamburg/ESO bright quasar survey
(Wisotzki et al.\ 2000), with the following additional criteria:
Right ascension between 9$^\mathrm{h}$ and 16$^\mathrm{h}$; 
nuclear absolute magnitudes $M_{B_J} < -23$; and
redshifts $z < 0.3$. The resulting sample consisted
of 30 targets distributed over 2200 deg$^2$, all of which were subsequently
observed. Because of the optical selection, the sample is predominantly radio-quiet,
containing only four bona fide radio-loud quasars.

\begin{figure}[tb]
\includegraphics[bb=114 137 400 400,clip,angle=0,width=10cm]{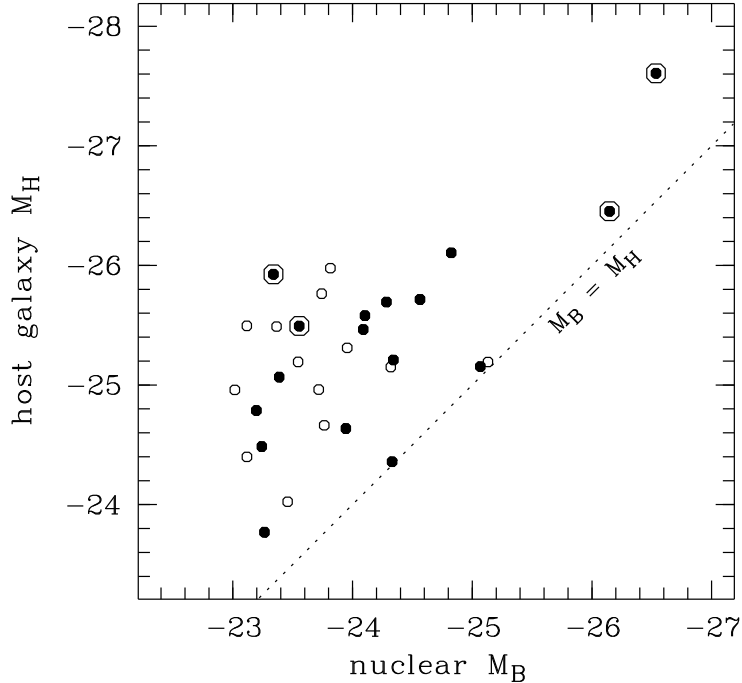}  
\caption{Host galaxy versus nuclear absolute magnitudes. Open symbols:
disk-dominated hosts; filled dots: elliptical hosts. The four radio-loud 
QSOs are marked as circled dots.}
\label{fig:mqmgal} 
\end{figure}

Observations were conducted in February 1999 with the ESO NTT and its
near-infrared camera SOFI. With a seeing between $0\farcs 5$ and $0\farcs 8$,
the host galaxy was clearly detected in all cases, and after PSF subtraction,
unique morphological type assignment was possible for most objects. 
A summary of the results:

\begin{itemize}
\item A disk model is preferred in 11 objects (37\,\%);
      there is often evidence for an additional substantial bulge component.
\item A spheroidal model is preferred in 16 objects (53\,\%); 
      for the QSOs with $M_{B_J} < -24$,
      10 out of 12 (83\,\%) are spheroidal.
\item Only three systems (10\,\%) are strongly interacting or irregular.
\item All hosts are very luminous, $M_H \la M_H^\star$ with $M_H \simeq -24.5$
      being a typical value for the general galaxy population (cf.\ below).
\item There are no hosts with $M_{\subs{$H$,\,gal}} > M_{\subs{$B$,\,nuc}}$,
      thus the McLeod et al.\ (1999) diagonal boundary is confirmed and 
      reinforced (in particular, there are no more upper limits).
\item Except for this boundary, there is no convincing evidence for 
      any correlation of $M_{\subs{$H$,\,gal}}$ with $M_{\subs{$B$,\,nuc}}$
      (apart from the two extreme outliers 3C~273 and PKS~1302$-$102, which
      however are flat-spectrum radio sources and probably beamed).
\end{itemize}

\section{Bivariate luminosity function}

In order to fully describe a QSO host galaxy luminosity function (QHGLF),
at least two independent variables are required: Host galaxy
luminosity $L_{\mathrm{gal}}$ and nuclear luminosity $L_{\mathrm{nuc}}$,
leading to a \emph{bivariate} luminosity function 
$\Phi (L_{\mathrm{gal}},L_{\mathrm{nuc}})$. Cosmological evolution 
of the QSO population demands $\Phi$ to depend additionaly on redshift $z$.

From the above results, we simplify the expression by tentatively assuming
that QSO and host galaxy LFs are formally uncorrelated over the luminosity
range of the sample under consideration.
$$
  \Phi \:=\: \phi(L_{\mathrm{nuc}},z)\:\psi(L_{\mathrm{gal}},z) .
$$
We adopt common parametric forms for both $\phi$ and $\psi$. The former
can be well approximated, for low redshifts and not too low luminosities,
by a single power law (K\"ohler et al.\ 1997; Wisotzki 2000), the latter
is usually expressed by a Schechter function. For the evolution we assume
a simple power-law form of pure number evolution, which is perfectly sufficient
at low redshifts. The resulting expression for the QHGLF is
$$
  \Phi \:=\: \Phi_0\,(1+z)^\kappa\,
             \left(\frac{L_{\mathrm{nuc}}}{L_{\mathrm{nuc}}^\star}\right)^\alpha\,
             \left(\frac{L_{\mathrm{gal}}}{L_{\mathrm{gal}}^\star}\right)^\beta\,
             e^{-L_{\mathrm{gal}}/L_{\mathrm{gal}}^\star}.
$$
An important additional feature is the 
McLeod et al.\ (1999) boundary, which basically states that 
for a given host galaxy mass (and hence luminosity), there is a well-defined
maximum nuclear luminosity, essentially given by the Eddington limit of
the central black hole.
Although the universal validity of this boundary has recently been questioned by
Percival et al.\ (2000; also these proceedings), it is clearly present in our data.
We therefore decided to incorporate it by
multiplying the above expression by a factor ${\cal H}(L_{\mathrm{gal}} - xL_{\mathrm{nuc}})$ 
where ${\cal H}(t_1-t_2) = 1$ for $t_1 > t_2$ and 0 elsewhere is the Heavyside step function,
and $x$ is an adjustable parameter. For $L_{\mathrm{gal}}$ measured in the $H$ band
and $L_{\mathrm{nuc}}$ measured in $B$, we adopt $x\approx 1$ (this corresponds to 
the diagonal dotted line in Fig.\ \ref{fig:mqmgal}).

\begin{figure}[tb]
\includegraphics[bb=114 137 455 400,clip,angle=0,width=10cm]{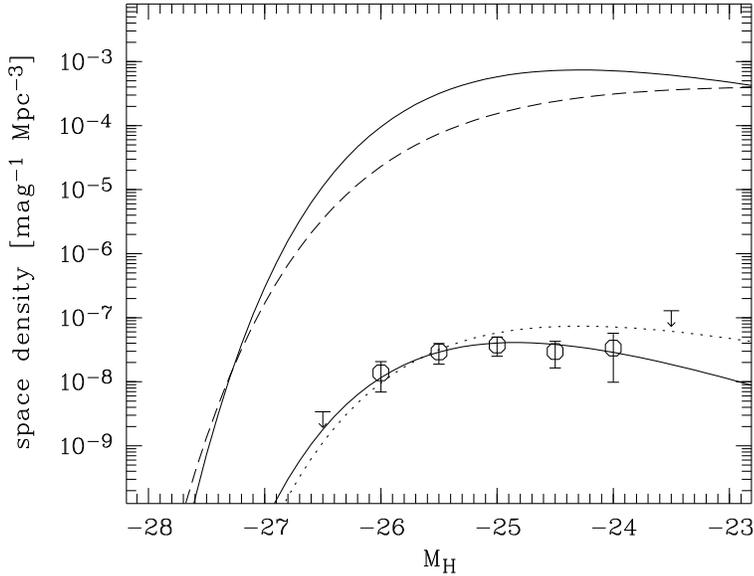}  
\caption{Estimates of the QSO host galaxy luminosity function. The points
give the binned QHGLF with errors and upper limits from Poisson statistics.
The lower solid curve is the best-fit maximum likelihood estimate.
Also shown are some LFs of early-type field galaxies for comparison. 
Upper solid curve: Lin et al.\ (1999); dashed curve: Kochanek et al.\ (2001). 
The dotted line is the Lin et al.\ LF with rescaled normalisation to 
approximately match the QSO data points.}
\label{fig:hglf} 
\end{figure}

\begin{figure}[tb]
\includegraphics[bb=114 137 400 400,clip,angle=0,width=10cm]{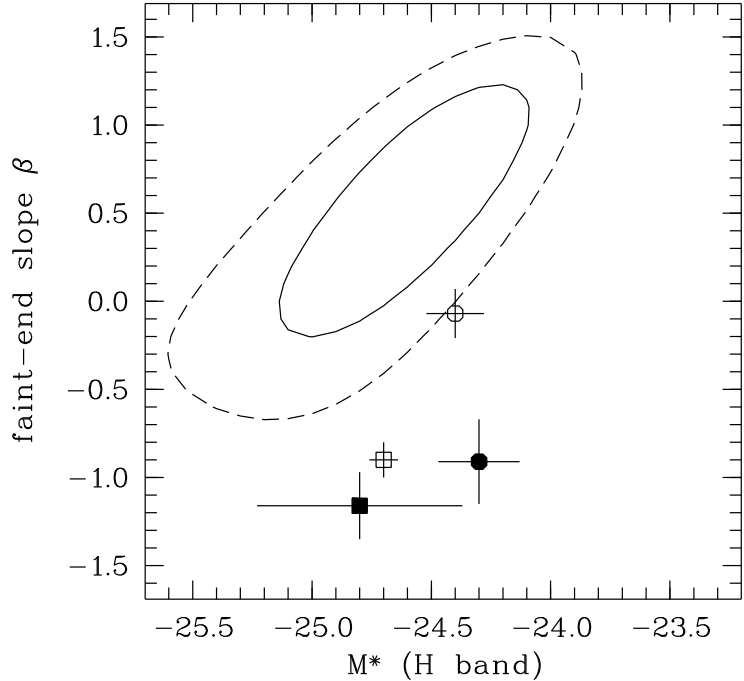}  
\caption{Likelihood contours for the HGLF parameters: The solid line corresponds
to 68\,\%, the dashed line to 90\,\% confidence. Comparison values for 
the field galaxy LF are given by symbol markers. Open circle: Lin et al.\ (1999);
filled circle: Gardner et al.\ (1997); open square: Kochanek et al. (2001);
filled square: Loveday et al.\ (2000).}
\label{fig:likelihood} 
\end{figure}

Numerical values of the QHGLF parameters have been estimated by maximum likelihood
fitting of the above functional form to the observed distribution of sources
in ($M_{\subs{$B$,\,nuc}},M_{\subs{$H$,\,gal}},z$) space. The resulting best-fit 
values are
\begin{eqnarray*}
  \Phi_0 & = & 1.3 \times 10^{-7}\:\mbox{Mpc}^{-3}\;\;\;\ \mathrm{for}\:\:M_{B,Q}^\star = -23 \\[-0.5ex]
  \kappa & = & 6.5 \\[-0.5ex]
  \alpha & = & -2.8 \\[-0.5ex]
  M_{H, \mathrm{gal}}^\star & = & -24.6 \pm 0.3 \\[-0.5ex]
  \beta  & = & 0.5 \pm 0.5
\end{eqnarray*}
We have also determined binned estimates of the QHGLF under the assumption that
the QSO luminosity function term $\phi(L_{\mathrm{nuc}},z)$ is well-described by
the above parameters. The resulting binned and maximum likelihood estimates
are shown in Fig.\ \ref{fig:hglf}.

The last two parameters are particularly interesting, as they describe the 
\emph{shape} of the galaxy luminosity function. In Fig.\ \ref{fig:likelihood} 
we show confidence contours for these parameters and compare
them with published values determined for the general field galaxy population
(adapted to the $H$ band by assuming typical galaxy colours).
The characteristic luminosity $M_{H, \mathrm{gal}}^\star$ is completely
within the range found for field galaxies, while the faint-end slope $\beta$ 
is only slightly flatter. The latter makes the QHGLF formally inconsistent 
with that of field galaxies, but Fig.\ \ref{fig:hglf} shows that the discrepancy
even at the faint end is not large: The dotted line represents the Lin et al.\ (1999) 
LF of early-type field galaxies, rescaled to approximately match the normalisation of
the QHGLF data points. The overall agreement is surprisingly good, and the 
slight mismatch at the faint end might even be explained by our limitation to 
only luminous QSOs.

To summarise, our results present no evidence that QSO host galaxies show
a luminosity distribution widely different from that of inactive field galaxies.
This lends considerable support to the hypothesis that the parent population
of quasars consists of ordinary early-type galaxies, with very little if any bias
towards higher luminosities or higher degree of morphological irregularities.

\section{Implications for the QSO duty cycle}

Dividing the space density of quasar hosts by that of the 
parent galaxies gives the ratio of active to inactive
galaxies as a function of galaxy luminosity. This ratio has been interpreted
in the past as the `probability' of finding a QSO in a galaxy.
By assuming that all major galaxies have massive black holes in their centres
and using the argument of `ensemble average equals time average',
this is furthermore the \emph{time fraction} that a typical galaxy spends in
a `quasar state' (also called the QSO duty cycle $\delta$).

From our determination of the QHGLF, we found that the parent population 
can be probably identified with normal early-type field galaxies. 
Under these premises, we can estimate $\delta$ directly from
the rescaling factor used in Fig.\ \ref{fig:hglf} to match the field galaxy
LF to the QSO data points, giving $\delta \sim 10^{-4}$.
Because of the similarity between the luminosity functions, there is virtually
no trend of $\delta$ with luminosity. We can confidently exclude that the number 
density ratio between quasar hosts and luminous field galaxies approaches unity 
(implying $\delta \simeq 1$) as suggested recently by Hamilton et al.\ (2001). 

Note that the above value for $\delta$ refers to zero redshift, since because of 
the explicit evolution term for the QSO space density in the analytic expression 
for $\Phi$, the QHGLF normalisation is taken at $z=0$. In this simple picture,
QSO evolution would be interpreted as an increase in the duty cycle towards
higher redshift with a rate of $\delta \propto (1+z)^{6.5}$.

With $0<z<0.3$ for our sample, this corresponds to a time span of
$T \sim 5$\,Gyrs. Multiplying this with the redshift-averaged value 
of $\delta(z)$, we can convert the duty cycle into a time scale of 
$<\!\Delta t\!> \simeq 10^6$ yrs, which is a robust lower limit to
the total time a typical low-redshift quasar is switched on, and 
probably not too different from the actual time scale for quasar-type
activity.

\end{document}